\documentclass[12pt]{article}
\usepackage{geometry}
\usepackage{xcolor}
\usepackage{a4}
\usepackage{alltt}
\usepackage{graphicx}
\usepackage{epsf}
\usepackage{amsmath}
\usepackage{amssymb}
\usepackage{cite}
\usepackage{hyperref}
\usepackage{blindtext}

\newcommand{\be}{\begin{equation}}
\newcommand{\ee}{\end{equation}}

\newcommand{\Rmnum}[1]{\expandafter\@slowromancap\romannumeral #1@}
\newcommand{\bea}{\begin{eqnarray}}
\newcommand{\eea}{\end{eqnarray}}
\begin{document}
\title{\bf Nonlinear anisotropy growth in Bianchi-I spacetime in metric $f(R)$ cosmology}
\author{Kaushik Bhattacharya, Saikat Chakraborty 
\thanks{ E-mail:~ kaushikb, snilch@iitk.ac.in} 
\\ 
\normalsize
Department of Physics, Indian Institute of Technology,\\ 
\normalsize
Kanpur 208016, India}
\maketitle 
\begin{abstract}
The present work is related to anisotropic cosmological evolution in
metric $f(R)$ theory of gravity.  The initial part of the paper
develops the general cosmological dynamics of homogeneous anisotropic
Bianchi-I spacetime in $f(R)$ cosmology. The anisotropic spacetime is
pervaded by a barotropic fluid which has isotropic pressure. The paper
predicts nonlinear growth of anisotropy in such spacetimes.  In the
later part of the paper we display the predictive power of the
nonlinear differential equation responsible for the cosmological anisotropy
growth in various relevant cases. We present the exact solutions
of anisotropy growth in Starobinsky inflation driven by quadratic
gravity and exponential gravity theory. Semi-analytical results are
presented for the contraction phase in quadratic gravity bounce. The
various examples of anisotropy growth in Bianchi-I model universe
shows the complex nature of the problem at hand.
\end{abstract}
\section{Introduction}

The issue related to stability of homogeneous and isotropic
cosmological solutions with respect to small anisotropy has been
studied intensely in theoretical cosmology \cite{Wainwright:1998ms},
\cite{Chen:2001fh}, \cite{Chen:2002ksa},
\cite{Barrow:2000ka}. Behavior of small anisotropy has been studied in
cosmological models, using general relativity (GR), in the contexts of
inflation \cite{Pereira:2015pga}, \cite{Anninos:1991ma},
\cite{Kitada:1992uh}, \cite{Kitada:1992uf}, \cite{Do:2017qyd} and
pre-bounce ekpyrotic contraction phase \cite{Garfinkle:2008ei},
\cite{Bozza:2009jx}, \cite{Barrow:2010rx}, \cite{Barrow:2015wfa}. In
the context of inflation the `No-Hair' conjecture asserts that any
pre-existing anisotropy must asymptotically die out in an inflating
universe. Wald has been able to prove the conjecture for all the
Bianchi models except Bianchi-IX \cite{Wald:1983ky} which requires a
large cosmological constant to isotropize the spacetime . In a
contracting universe, provided the universe is dominated by a matter
component mimicking an ultra stiff barotropic fluid, growth of small
anisotropy is suppressed with respect to that of the Hubble
parameter. In absence of any such fluid in a contracting phase, small
anisotropy grows large and dominates over all other matter
components. This leads to the Belinsky-Khalatnikov-Lifshitz (BKL)
instability \cite{Belinsky:1970ew}, foiling the
bounce. Mathematically, it can be shown that in presence of a slowly
rolling scalar field the isotropic de-Sitter solution is an attractor
for an expanding universe and in presence of a fast rolling scalar
field the isotropic power law solution (for the scale-factor) is an
attractor for contracting universe. Therefore an inflationary scenario
is usually realized by a slowly rolling scalar field and an ekpyrotic
scenario is usually realized by a fast rolling scalar field
\cite{Sharma:2015hka}, \cite{Panda:2015wya}, \cite{Solomons:2001ef},
\cite{Cai:2013kja}, \cite{Cai:2014bea}.

In the present work we have analyzed the evolution of spacetime
anisotropy in $f(R)$ gravity\footnote{For a general understanding of
  modified gravity theories one can look at the review in
  Ref.~\cite{Nojiri:2017ncd}.}, where the analysis becomes
significantly more involved than that of models based on GR.  Some
attempts in modified, quadratic gravity\cite{Muller:2017nxg,
  Toporensky:2016kss} do discuss about anisotropic cosmologies while
analyzing the past stages of a cosmological system near the
singularity.  In these works the authors show that near the
singularity the universe may have an anisotropic mode of
existence. The papers in general do not address a cosmological bounce
scenario.  Although previously there have been some progress in
generalizing the no-hair theorem to incorporate higher order gravity
theories \cite{Maeda:1987xf}, \cite{Cotsakis:1993er},
\cite{Schmidt:1988hq}, \cite{Starobinsky:1980te} and some applications
of dynamical system analysis to understand anisotropic cosmology in
higher order gravity \cite{Leon:2014dea}, \cite{Leon:2010pu},
\cite{DeFelice:2013awa}, the previous attempts missed an important
property of anisotropic cosmological dynamics related to nonlinear
growth of anisotropy in the homogeneous and anisotropic Bianchi-I type
of spacetime. In the present paper we first show analytically that in
Starobinsky inflation any initial anisotropy will rapidly fade away.

It is then shown that  in contraction phases in Bianchi-I metric where
specifying  an unique  scale-factor  for contraction  does not  always
yield a  unique cosmological development.  This result is  possible in
$f(R)$  cosmology  and in  GR  one  cannot  have this  property.  This
non-uniqueness of cosmological development corresponding to a specific
scale-factor opens up a  new problem as cosmological evolution becomes
more complex conceptually and as a consequence only simple $f(R)$ models
can be  semi-analytically solved. Any  $f(R)$ model which is  a higher
order  polynomial  in  $R$  compared  to the  quadratic  $f(R)$  model
requires a complete numerical solution for anisotropy growth.  We show
our  result  in  quadratic  $f(R)$  model,  which  is  gravitationally
unstable if it has to  accommodate a cosmological bounce. Our work may
be taken as an effective toy model which is used to crack a formidable
problem  in  cosmological dynamics.  To  our  understanding the  above
mentioned topics are  discussed for the first time  in full generality
in the present paper.

We also present some preliminary results of non-linear anisotropy
growth in exponential gravity models. The exponential gravity model
has two exact solutions. One exact solution is a bouncing solution in
presence of matter and the other exact solution is a expanding
universe solution at a de-Sitter point. In these cases the anisotropy
generation equation turns out to be a transcendental equation. We
present some simple solutions in this case probing the nature of
growth of small initial anisotropy. 

The material in the paper is organized in the following way. The
second section discusses about the basics of Bianchi-I spacetimes and
sets the notations and conventions followed throughout the paper. In
section \ref{formu} we present the general formalism of homogeneous
and anisotropic cosmological dynamics in metric $f(R)$ cosmology. This
part contains important results.  In this section for the first time
one comes across the complex nature of anisotropy development. In
section \ref{chroots} we present the results related to nonlinear
anisotropy growth in quadratic $f(R)$ theory induced inflation and
cosmological bounce. Section \ref{expgrv} presents the results of
anisotropy growth for some exact results in exponential gravity. The
next section is the concluding section where we summarize the results
obtained in the paper.
\section{The anisotropic Bianchi-I metric and its properties}
\label{one}

For our analysis, we have used the metric for Bianchi-I spacetime,
\begin{equation}
ds^2=-dt^2+a_1^2(t)dx_1^2+a_2^2(t)dx_2^2+a_3^2(t)dx_3^2\,,
\label{b1}
\end{equation}
where $a_1(t)$, $a_2(t)$ and $a_3(t)$ are the different scale-factors,
whose relative differences specify the amount of anisotropy in the
evolving universe. Existence of such anisotropic cosmological models
in higher order gravity theories have been extensively studied in
literature \cite{Middleton:2010bv}, \cite{Shamir:2017aon},
\cite{Clifton:2006kc}, \cite{Banik:2016qpq}.  In most of the earlier
attempts the authors have tried to find out the nature of anisotropic
spacetimes using various forms of anisotropic metric and using various
forms of gravitational Lagrangians. In the present paper we show that
the previous attempts have missed a vital ingredient in anisotropic
expansion/contraction. The effect we discuss is clearly visible in
Bianchi-I spacetime, but we think similar effects may be present in
other anisotropic cosmological models.

In this paper we will assume the presence of a perfect hydrodynamic
fluid, in the Bianchi type-I spacetime, whose energy-momentum tensor
(EMT)is given by $T_{\mu \nu} = (\rho + P)u_\mu u_\nu + P g_{\mu
  \nu}$, where $\rho$ is the energy-density and $P$ specifies
isotropic pressure of the perfect fluid. The 4-velocity of the fluid
element is given by $u_\mu$, which being a time-like vector is
normalized as $u^\mu u_\mu = -1\,.$ Although the spacetime metric is
anisotropic the fluid which pervades the spacetime is assumed to be
isotropic. In this paper we will assume a barotropic equation of state
for the perfect fluid, $P=\omega \rho\,,$ where $\omega$ specifies the
barotropic ratio.

One can rewrite the form of the anisotropic metric, given in
Eq.~(\ref{b1}), in terms of the (geometric) average of the three
scale-factors given by $a(t)=\left[a_1(t)\,a_2(t)\,a_3\right]^{1/3}$.
The three different scale-factors in terms of the geometric average
scale-factor can be written as, $a_i(t)=a(t)e^{\beta_i(t)}$, where
$i=1\,,2\,,3$\footnote{Latin alphabets as $i$ and $j$ run from 1 to 3
  where as Greek alphabets run from 0 to 3.}. The time dependent
functions $\beta_i(t)$ specify the anisotropy in the metric and they
are constrained as
\begin{eqnarray}
\beta_1+\beta_2+\beta_3=0\,.
\label{betar}
\end{eqnarray}
Using the above relations one can now rewrite the metric given in
Eq.~(\ref{b1}) as
\begin{eqnarray}
ds^2=-dt^2+a^2(t)\left[e^{2\beta_1(t)}dx_1^2 + e^{2\beta_2(t)}dx_2^2 +
e^{2\beta_3(t)}dx_3^2\right]\,.
\label{b1m}
\end{eqnarray}
In this notation one can define the Hubble
parameter, as an arithmetic average, and its time-derivative as
\begin{eqnarray}
H(t)\equiv \frac{1}{3}\left(\frac{\dot{a}_1}{a_1}+\frac{\dot{a}_2}{a_2}+
\frac{\dot{a}_3}{a_3}\right)=\frac{\dot{a}}{a}\,,
\,\,\,\,
\dot{H}(t)=\frac{\ddot{a}}{a}-\frac{\dot{a}^2}
{a^2}\,.
\label{hb2}
\end{eqnarray}
Mainly for the sake of brevity, henceforth in this article we will
omit the word average (either geometric or arithmetic) before
scale-factor or Hubble parameter. In presence of anisotropy the Ricci
scalar turns out to be
\begin{equation}
R=6(\dot{H}+2H^2)+(\dot{\beta_1}^2+\dot{\beta_2}^2+\dot{\beta_3}^2)\,.
\label{R_bianchi1}
\end{equation}
In the next section we formulate the anisotropic cosmological dynamics
guided by metric $f(R)$ theory. The fact that the derivatives of the
anisotropy parameters are themselves present in the expression of the
Ricci scalar will make anisotropic cosmological dynamics much more
involved in $f(R)$ gravity, compared to the general relativistic case. 
\section{Formulation of anisotropic cosmological dynamics guided by metric
  $f(R)$ theory}
\label{formu}

The field equation in $f(R)$ gravity is
\begin{eqnarray}
G^\mu_{\,\,\nu}=\frac{\kappa}{f^\prime(R)}[T^\mu_{\,\,\nu} +
  T^\mu_{\,\,\nu\,(\rm{curv})}]\,,
\label{frfe}
\end{eqnarray}
where $G^\mu_{\,\,\nu}$ is the Einstein tensor and
$T^\mu_{\,\,\nu\,(\rm{curv})}$ is the energy momentum tensor due to
curvature. Here $\kappa=8 \pi G$, where $G$ is the Newton's
gravitational constant and is related to the Planck mass $M_P$ via
$G=1/M_P^2$. In the present paper we will approximately use $M_P
\approx 10^{19}\,{\rm GeV}$. The prime on top right hand side of any
function represents the ordinary derivative of that function with
respect to the Ricci scalar $R$. In particular
\begin{eqnarray}
 T^\mu_{\,\,\nu\,(\rm{curv})} \equiv \frac{1}{\kappa}
\left[-\left(\frac{Rf^\prime(R)-f(R)}{2} +
\Box f^\prime(R)\right)\delta^\mu_\nu + g^{\mu \alpha} D_\alpha D_\nu 
f^\prime(R)\right]
\label{tmunucurv}
\end{eqnarray}
where $D_\mu A_\nu$ is the covariant derivative of the covariant
4-vector $A_\nu$ and $\Box \equiv g^{\alpha \beta} D_\alpha D_\beta$.
The $0-0$ component of the field equation in $f(R)$ theory in an
anisotropic spacetime is then given as, $G^0_{\,\,0} =
-\frac{\kappa}{f^{\prime}(R)}\left(\rho+\rho_{\rm{curv}}\right)$,
where 
\begin{eqnarray}
\rho_{\rm{curv}} = \frac{1}{\kappa}\left[\frac{Rf^{\prime}(R)-f(R)}{2}-
3H\dot{R}f^{\prime\prime}(R)\right]\,. 
\label{rho_curv}
\end{eqnarray}
The other three equations, specifying the $i-j$ components become, 
$G^i_{\,\,j} =
\frac{\kappa}{f^{\prime}(R)}\left(T^i_{\,\,j}+T^i_{\,\,j\,(\rm{curv})}
\right)$, where $i,j=1\,,\,2\,,\,3$. Here $T^i_{\,\,j}=P\delta^i_j$ stands for
pressure of the perfect hydrodynamic fluid(s) whose EMT(s) has(have)
the same form as specified in section \ref{one}.  The form of
$T^i_{\,\,j\,(\rm{curv})}$ is given as
\begin{eqnarray}
\kappa T^i_{\,\,j\,(\rm{curv})} &=& -\left[\frac{Rf^\prime(R) -f(R)}{2} 
-\ddot{R}f^{\prime \prime}(R) - \dot{R}^2 f^{\prime \prime \prime}(R) - 
2 H\dot{R}f^{\prime \prime}(R) \right]\delta^i_j\nonumber\\
&& - B^i_{\,\,j}\dot{R}f^{\prime \prime}(R)\,,
\label{inttii2}
\end{eqnarray}
where the components of the tensor $ B^i_{\,\,j}$ are defined as
\begin{eqnarray}
B^1_{\,\,1} = \dot{\beta}_1\,,\,\,\,
B^2_{\,\,2} = \dot{\beta}_2\,,\,\,\,
B^3_{\,\,3} = \dot{\beta}_3\,,\,\,\,
B^i_{\,\,j} = 0\,, {\rm if}\,,\,\,i\ne j\,.
\label{B}
\end{eqnarray}
In terms of the above quantities one can now write,
\begin{eqnarray}
G^i_{\,\,j} =
\frac{\kappa}{f^{\prime}(R)}\left(P + P_{\rm{curv}}
\right)\delta^i_j - B^i_{\,\,j}\dot{R}f^{\prime \prime}(R)\,,
\label{giifr}
\end{eqnarray}
where
\begin{eqnarray}
P_{\rm{curv}} &=& \frac{\dot{R}^2f^{\prime\prime\prime}+2H\dot{R}
f^{\prime\prime}+\ddot{R}f^{\prime\prime}}{\kappa}-\frac{Rf^{\prime}-f}{2\kappa}\,.
\label{p_curv}
\end{eqnarray}
In terms of the Hubble parameter and the anisotropy parameter,
the $0-0$ component of the field equation becomes
\begin{eqnarray}
H^2 = \frac{\kappa}{3f^{\prime}(R)}
\left(\rho+\rho_{\rm{curv}}\right) + \frac{1}{6}\sum_{i=1}^3
\dot{\beta}_i^2\,,
\label{hsquare}
\end{eqnarray}
while Eq.~(\ref{giifr}) becomes
\begin{eqnarray}
2\dot{H}+3H^2-3H\dot{\beta}_i-\ddot{\beta}_i+\frac{1}{2}
\sum_{i=1}^3 \dot{\beta}_i^2
= -\frac{\kappa}{f^{\prime}(R)}\left(P + P_{\rm{curv}}
\right) + \dot{\beta}_i\dot{R}\frac{f^{\prime \prime}(R)}{f^\prime(R)}\,.
\label{hdoteqn}
\end{eqnarray}
Adding the three equations, corresponding to each value of the index
$i$ in the above expression, one gets
\begin{eqnarray}
2\dot{H}+3H^2 
= -\frac{\kappa}{f^{\prime}(R)}\left(P + P_{\rm{curv}}
\right) - \frac{1}{2}
\sum_{i=1}^3 \dot{\beta}_i^2\,.
\label{hdot2}
\end{eqnarray}
If one uses Eq.~(\ref{hsquare}) in the above equation then one gets
$\dot{H}$ as
\begin{eqnarray}
\dot{H}
= -\frac{\kappa}{2f^{\prime}(R)}\left[(1+\omega)\rho + (\rho_{\rm
    curv} + P_{\rm{curv}})\right] - \frac{1}{2}
\sum_{i=1}^3 \dot{\beta}_i^2\,,
\label{hdot3}
\end{eqnarray}
where $P=\omega \rho$ has been used. In the present case we define the
anisotropy factor $x$ as
\begin{eqnarray}
x^2(t)\equiv \sum_{i=1}^3 \dot{\beta}_i^2(t)\,.
\label{bt}
\end{eqnarray}
For an isotropic universe $x^2=0$ for all values of $t$, implying that
all the $\beta_i$'s are constant in time.  In such a case one can
appropriately make (time-independent) coordinate rescaling in an
appropriate way to make the spacetime look exactly like the
Friedmann-Lemaitre-Robertson-Walker (FLRW) spacetime.  Using
Eq.~(\ref{betar}) and Eq.~(\ref{hdoteqn}) and the above definition of
the anisotropy factor one can show that $x$ satisfies the differential
equation:
\begin{eqnarray}
\dot{x} + \left(3H + \frac{\dot{f}^{\prime
  }(R)}{f^\prime(R)}\right)x=0\,,
\label{ddotbeqn}
\end{eqnarray}
whose (nontrivial) solution must be like
\begin{eqnarray}
x=\frac{b}{a^3(t) f^\prime(R)}\,,
\label{dotb}
\end{eqnarray}
where $b$ is a real integration constant. The above equation contains
the most important theoretical ingredient of the present paper. The
Ricci scalar in the present case can be written as $R=6(\dot{H}+2H^2)+
x^2$ which depends on $x$ and the last equation shows $x$ is a
function of $R$ in $f(R)$ gravity. As a consequence of the above
relation in $f(R)$ gravity, one cannot define an unique anisotropy
dynamics. For any given $f(R)$ gravity, in general multiple time
evolutions of the anisotropy factor $x$ is possible, each
corresponding to a different equation of state for the barotropic
fluid. In the present case  the  time derivative of the Ricci scalar is
\begin{eqnarray}
\dot{R}
= \frac{6(\ddot{H}+4H\dot{H}-Hx^2)}{\left(1 +  2\frac{f^{\prime \prime}(R)}
{f^\prime(R)}x^2\right)} \,. 
\label{rbd}
\end{eqnarray}
Working out similarly one can write,
\begin{eqnarray}
\ddot{R}&=&
\frac{6(\dddot{H}+4H\ddot{H}+4\dot{H}^2)  
-2\left[\left(3\dot{H} + \frac{\dot{R}^2 
f^{\prime\prime\prime}}{f^\prime}- \frac{\dot{f}^{\prime \,2
  }}{f^{\prime \,2}}\right)
- 2\left(3H + \frac{\dot{f}^{\prime
}}{f^\prime}\right)^2\right] x^2}{\left(1 +2x^2\frac{f^{\prime\prime}}{f^\prime}\right)}\,.
\nonumber\\
\label{rbdd}
\end{eqnarray}
The above equations show that once we know the form of $x$ in terms of
the scale-factor, we can write the values of $R$, $\dot{R}$ and
$\ddot{R}$ in terms of the scale-factor. The cosmological
dynamics of anisotropic $f(R)$ theory is encoded in
Eq.~(\ref{hdot3}), Eq.~(\ref{dotb}) and the
energy-momentum conservation equation
\begin{eqnarray}
\dot{\rho} + 3 H \rho(1+\omega)=0\,.
\label{rhod}
\end{eqnarray}
The specification of $\omega$ and this set of three equations and the
initial conditions specifying $a$, $\dot{a}$, $\ddot{a}$, $\dddot{a}$,
$b$ and initial $\rho$ are enough to specify the anisotropic dynamics
in $f(R)$ cosmology if Eq.~(\ref{dotb}) has an unique root.  If
Eq.~(\ref{dotb}) does not have an unique root then the initial
conditions must have to be enhanced.  In the next section we will show
when the above list of initial conditions require to be enhanced.

The first order differential equation in Eq.~(\ref{ddotbeqn})
specifies the growth of anisotropy in metric $f(R)$ gravity models
where spacetime is specified by a Bianchi-I model.  From the form of
the equation it is seen that the amount of nonlinearity in
Eq.~(\ref{ddotbeqn}) depends upon time.  It can be noted that $x=0$
have some interesting properties. The first thing to note about this
point is that at $x=0$ one always has $\dot{x}=0$ .  The other
interesting properties about this point are as follows.
\begin{enumerate}
\item If the system resides at the point $x=0$ then is is impossible
  to perturb the system to have non-zero values of $x$. The system can
  have $x=0$ value only when $b=0$, and $b$ is specified by the initial
  condition. Consequently if the initial condition is such that $x=0$
  then there will be no anisotropy growth in the future.

\item On the other hand if the initial condition is such that $b \ne
  0$ then the system will never reach $x=0$ unless $a^3(t)f^\prime(R)$
  diverges in finite time, signifying a cosmological singularity.
\end{enumerate}

As $x$ cannot be zero in the future if $b\ne 0$ in a non-singular
cosmology, the important parameter which keeps track of effective
anisotropy is given by the factor $x^2/H^2$. From Eq.~(\ref{hsquare}),
Eq.~(\ref{hdot2}) and the expression of the Ricci scalar, $R$, it can
be verified that when $x^2/H^2 \ll 1$ one can safely neglect the
effect of anisotropy in the cosmological dynamics of Bianchi-I type
models.
\section{Evolution of the anisotropic factor $x(t)$ in quadratic gravity}
\label{chroots}

In this section we will focus on quadratic gravity 
\begin{eqnarray}
f(R)=R + \alpha R^2\,,
\label{quadfr}  
\end{eqnarray}
where $\alpha$ is a real number. Although this is a simple form of
$f(R)$ but it can be used to model cosmological inflation as well as
cosmological bounce for positive and negative signs of the constant
$\alpha$ respectively\footnote{For bounce one must have $\alpha<0$ as
  shown in \cite{Paul:2014cxa}}. In this section we will determine the
evolution of $x(t)$ in quadratic gravity. The technique of evolution
of $x(t)$ in higher order gravity will be similar but much more
involved. For higher order polynomial gravity the order of the
algebraic equation yielding the roots of $x(t)$ may be five (or
higher) and consequently there does not exist any general algebraic
formalism yielding those roots.

From Eq.~(\ref{dotb}) one can easily verify that the algebraic
equation specifying $x(t)$ in quadratic gravity is a cubic equation of
the form:
\begin{eqnarray}
x^3 + A_1 x +A_2=0\,,
\label{cubice}
\end{eqnarray}
where
\begin{eqnarray} 
A_1 &=& 6 (\dot{H}+2H^2) + \frac{1}{2\alpha}\,,
\label{a1}\\
A_2 &=& -\frac{b}{2 \alpha a^3}\,.
\label{a2}
\end{eqnarray}
The discriminant, $\Delta$, specifying the roots and their properties is given by
\begin{eqnarray}
\Delta = -4 A_1^3 -27A_2^2\,. 
\label{discrim}
\end{eqnarray}
If $\Delta>0$ there will be three distinct real roots, if $\Delta<0$
then there will be one real root (and two complex roots) and if
$\Delta=0$ there can be repeated real roots. The roots of
Eq.~(\ref{cubice}) are as follows:
\begin{eqnarray}
  x=\left\{
  \begin{array}{c}
  -\frac{(2/3)^{1/3}A_1}{(-9A_2 + \sqrt{-3\Delta})^{1/3}}
    + \frac{(-9A_2 + \sqrt{-3\Delta})^{1/3}}{2^{1/3}\,3^{2/3}}\,,\\
  \frac{(1\pm i\sqrt{3})A_1}{2^{2/3}\,3^{1/3}\,(-9A_2 + \sqrt{-3\Delta})^{1/3}}
   - \frac{(1\mp i\sqrt{3})(-9A_2 + \sqrt{-3\Delta})^{1/3}}{2^{4/3}\,3^{2/3}}\,. 
  \end{array}
\right.
\label{croots}
\end{eqnarray}
In terms of trigonometric functions the above roots can be represented
as,
\begin{eqnarray}
  x=\left\{
  \begin{array}{c}
  2\sqrt{-\frac{A_1}{3}}\cos\left[\frac{1}{3}\tan^{-1}\left(\frac{\sqrt{3\Delta}}
{-9A_2}\right)\right]\,,\\
  -2\sqrt{-\frac{A_1}{3}}\cos\left[\frac{1}{3}\left(\pi\mp\tan^{-1}
\left(\frac{\sqrt{3\Delta}}{-9A_2}\right)\right)\right]\,. 
  \end{array}\right.
\label{trigroot}
\end{eqnarray}
In the numerical calculations we will use the above form of the roots
as they are less cumbersome to handle when all the roots are real. If
initially the three roots of Eq.~(\ref{cubice}) are all real then one
does require a separate initial condition, specifying a particular
initial root out of the three possible roots, to describe the
anisotropic cosmological dynamics in $f(R)$ gravity. On the other hand
if there is only one real root then the added initial condition looses
its significance and the initial conditions as specified in the last
section is enough to describe the cosmological dynamics.

From Eq.~(\ref{dotb}) it is seen that the anisotropy factor depends
upon $a$, $\dot{a}$, $\ddot{a}$ and the integration constant $b$. More
over from the form of the cubic equation followed by $x$ it can be
easily seen that out of the three roots one tends to vanishes when $b
\rightarrow 0$, where as the other two roots in general do not tend to
zero when $b$ becomes arbitrarily small. If all the roots are real
then the root which vanishes when $b$ vanishes plays an important role
as in this case one can tune the value of the initial anisotropy by
tuning the value of $b$.  When the system admits only one real root
then this root always tends to zero when $b$ tends to zero. In GR,
when one deals with anisotropic Bianchi Type-I cosmology, the equation
followed by the anisotropy factor is $x(t)=b/a^3(t)$ and hence no such
complications arise.
\subsection{Starobinsky inflation}

We can now apply our formalism to get the first nontrivial result
related to anisotropic cosmological dynamics in Starobinsky's model of
inflation. In this model of inflation the universe inflates in absence
of any hydrodynamic fluid.  In quadratic gravity inflation the
parameter $\alpha$ appearing in Eq.~(\ref{quadfr}) is always positive
which makes $f^\prime \equiv df/dR >1$. In presence of anisotropy the
inflating spacetime shows very fast growth in the average scale-factor
$a(t)$ while the average Hubble parameter satisfies the condition
$\dot{H}=-\epsilon H^2$ where $\epsilon$ is a slow-roll parameter.
During inflation $\epsilon \ll 1$ and this condition prevails until
$\epsilon \sim 1$ at the end of inflation\cite{DeFelice:2010aj}. In
the excellent review on Starobinsky inflation given in
Ref.\cite{DeFelice:2010aj} the authors use the slow-roll approximation
to derive the properties of the inflating FLRW spacetime. In this
paper we will use the conventions of the above reference but will not
exactly apply slow-roll mechanism. We will use the full $f(R)$ theory
equations as discussed in the last section with specific inflationary
initial condition which gives rise to a rapidly expanding universe. In
a later publication we want to generalize slow-roll conditions in
quadratic gravity inflation in anisotropic Bianchi-I spacetimes.

In this subsection we show that any kind of anisotropy, if present
initially, will be damped during the inflationary phase in quadratic
gravity. We analytically prove our result for large initial
anisotropy, the proof remains the same for small initial
anisotropy. If initially the anisotropy factor was large then
inflation will successfully isotropize the universe and there will be
no remaining anisotropy at the end of inflation. Although this fact
is known to be true in inflationary models based on GR\cite{Wald:1983ky}, in
this article we show that similar outcome is also expected in
quadratic theory of inflation. First we show that the maximum
anisotropy allowed in quadratic gravity, during inflation, has a
maximum bound:
\begin{eqnarray}
x \le \sqrt{6}\,H\,, 
\label{anisb}
\end{eqnarray}
consequently the maximum anisotropy which can be isotropized is
related with the Hubble parameter. To prove the above assumption one
must note that in quadratic gravity inflation, $\rho=0$, and
the inflationary phase is initiated by curvature energy density $\rho_{\rm curv}$.
The anisotropy energy contribution, in Eq.~(\ref{hsquare}), is
non-negative and consequently for inflation to start initially (when
ideally the anisotropy effect is maximum) $\rho_{\rm{curv}}>0$. The
fact that the curvature energy-density appearing in the constraint
Eq.~(\ref{hsquare}) as:
$$H^2 = \frac{\kappa \rho_{\rm{curv}}}{3f^{\prime}(R)} + \frac{x^2}{6}\,,$$
cannot be negative during inflation justifies Eq.~(\ref{anisb}).

Even if the initial anisotropy present in the universe is given by the
maximum bound of $x$ in Eq.~(\ref{anisb}) the anisotropy factor
rapidly fades away during quadratic inflation. To show this we first
note that in Starobinsky inflation $A_1>0$, as $|\dot{H}|\ll H^2$. As
a result the discriminant $\Delta <0$ implying that there is only one
unique real root of Eq.(\ref{cubice}). This root is given by the top
right hand side term in Eq.~(\ref{croots}). From the expression of
$A_2$ one can see that it rapidly diminishes in an inflating universe
and one can assume $A_1^3 > A_2^2$ after some time from the onset of
inflation. Assuming that $A_2 \to 0$ rapidly after inflation starts one can
easily show that the relevant real root of Eq.~(\ref{anisb}) tends to
zero during inflation.
\begin{figure}[!t]
\centering
\includegraphics[scale=.5]{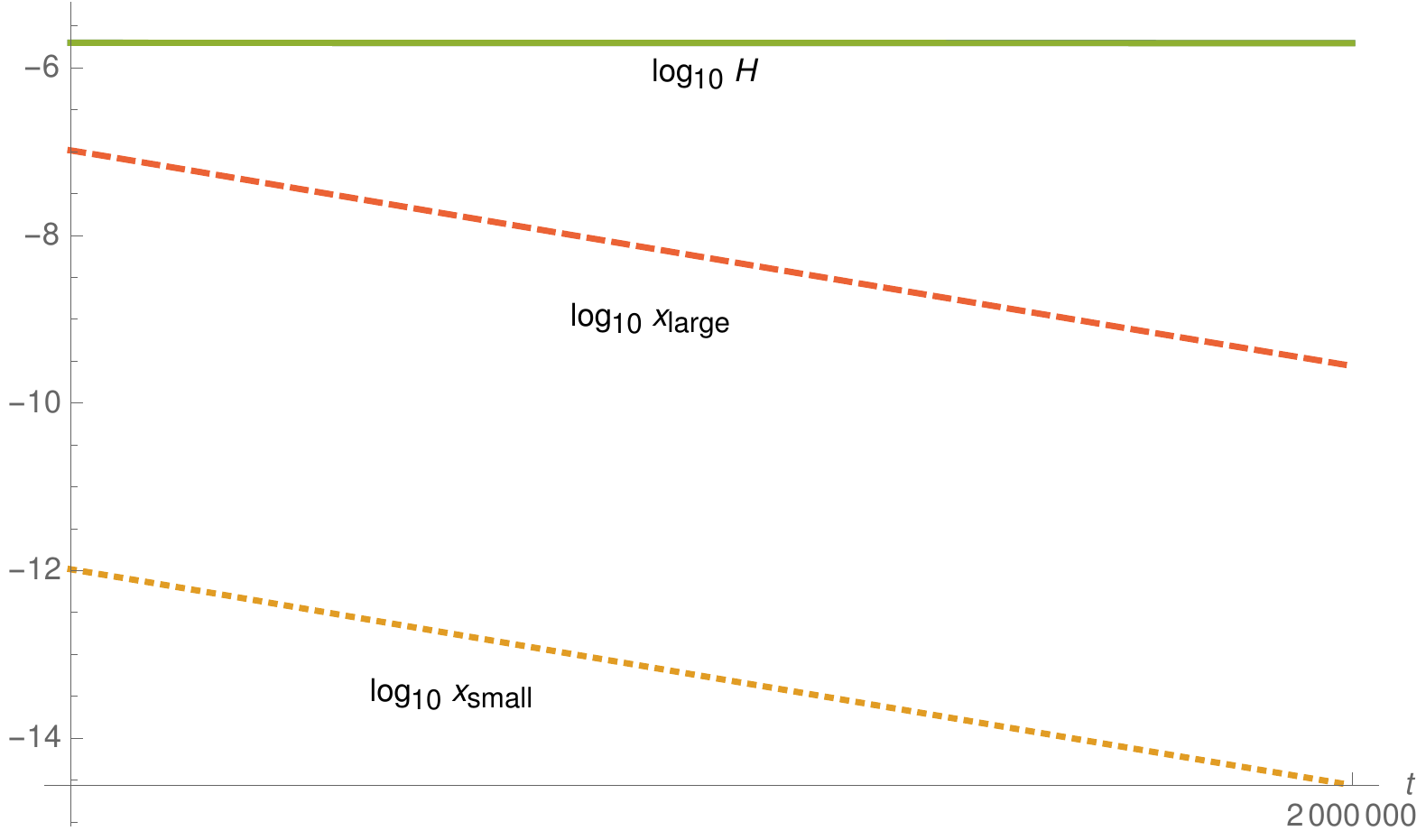}
\caption{Plot showing evolution of $H$ and $x$ in logarithmic scale
  during inflationary phase. Relatively large initial anisotropy
  $x_{\rm large}$ corresponding to $b=10^{-5}$ is shown by the dashed
  line.  Small initial anisotropy $x_{\rm small}$ corresponding to
  $b=10^{-10}$ is plotted by the dotted line. The Hubble parameter
  remains approximately the same in both the cases. Time axis spans
  from $10^6$ to $2\times 10^6$ and initial $H\sim 2\times 10^{-6}$
  (for both the cases of large and small initial anisotropy). Planck
  units are used for time and $H$.}
\label{fig1}
\end{figure}

We will now present a numerical solution of the cosmological dynamics
during inflation and point out that the anisotropy factor $x^2/H^2$
always remains sufficiently smaller and at no stages of quadratic
inflation $x^2 \approx H^2$. Here we assume that inflation starts at,
$t_i=10^6$, in Planck units. In this unit system the actual value of a
quantity is obtained by multiplying the value of the physical quantity
by a particular power of Plank mass $M_{P}$. The specific power
corresponds to the mass dimension of the physical quantity. In the
present case the actual value of $t_i$ is $t_i \,M^{-1}_{P}$.  In this
article we will assume $M_P \approx 10^{19}\,{\rm GeV}$ and
consequently $t_i = 10^{-13}\,{\rm GeV}^{-1}$ expressed in energy
units. Expressed in the seconds $t_i \approx 10^{-37}s$ and that is
$10^6$ times Planck time expressed in seconds.  The value of $\alpha$
is chosen as $\alpha=10^{12}$ which in natural units will be $10^{12}
M_P^{-2}$.  Phenomenologically one expects quadratic correction to
Einstein gravity at a very early phase of the universe when $H\sim
10^{12-13}$GeV or more. For this benchmark value of the $H$ the Ricci
scalar $R \sim 10^{26}\,{\rm GeV}^2$ assuming $x^2 < H^2$. If the
quadratic correction $\alpha R^2$ becomes effective at such a value of
$R$ then $\alpha \sim 1/R$ yielding $\alpha \sim 10^{-26}\,{\rm
  GeV}^{-2} = 10^{12}\,M_P^{-2}$, justifying our choice of $\alpha$.
\begin{figure}[!t]
\centering
\includegraphics[scale=.6]{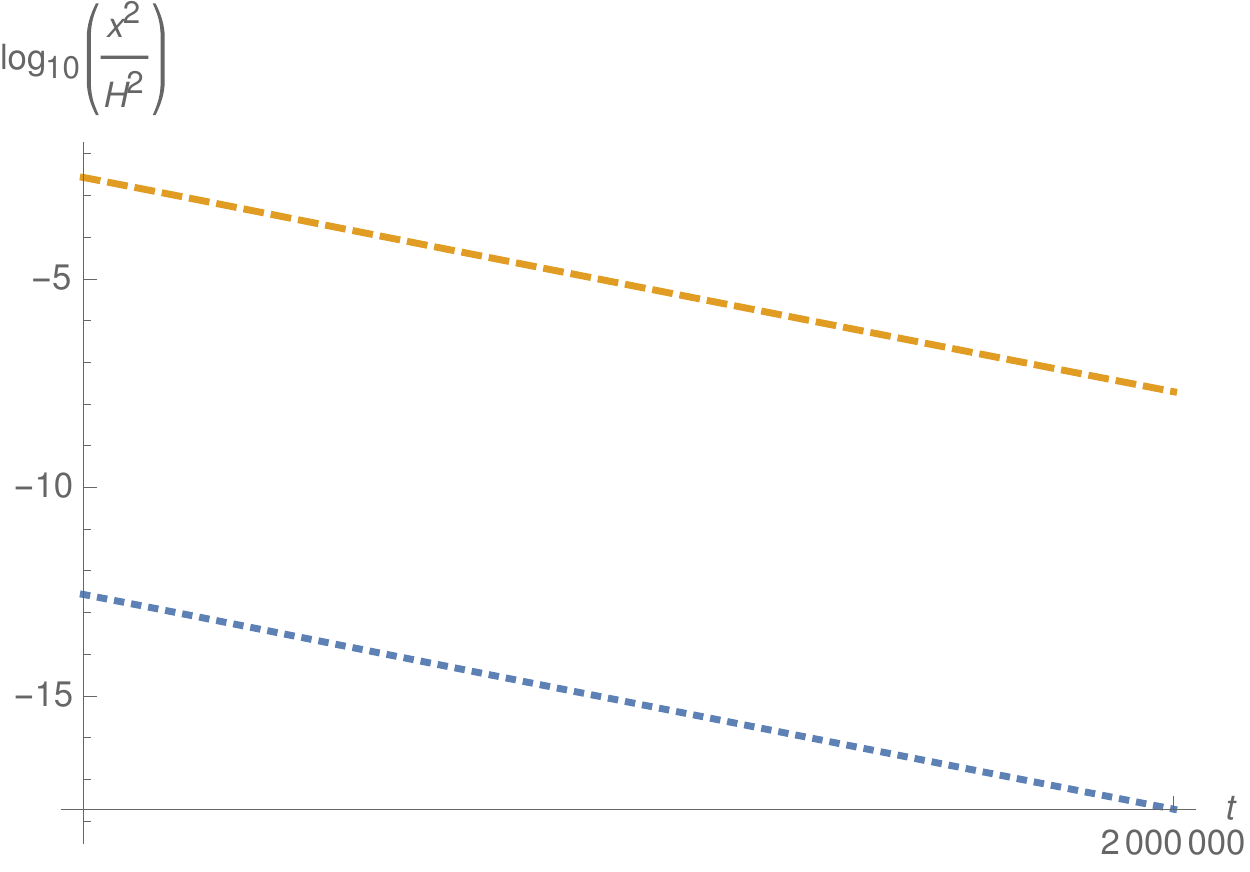}
\caption{Evolution of the $x^2/H^2$ in logarithmic scale for the cases
  corresponding to relatively large anisotropy where $b=10^{-5}$, 
  in dashed curve and relatively small anisotropy, where
  $b=10^{-10}$, in dotted curve. Time span and initial Hubble parameter value
  remains the same as specified in the caption of Fig.~\ref{fig1}.}
\label{fig2}
\end{figure}
The initial Hubble parameter is chosen to be $H(t_i)=2 \times 10^{-6}$
in Planck units, its value in standard units is $10^{13}\,{\rm
  GeV}$. Inflation ends at $t_f \approx t_i + 70\times H(t_i)^{-1} =
36 \times 10^6$, which corresponds to slightly more than 70
e-folds. The consistent inflationary\footnote{Here we apply the
  initial conditions for inflation on the average scale-factor and
  Hubble parameter and its derivatives. By consistent initial
  conditions we mean that these initial conditions satisfy all the
  constraints of quadratic gravity inflation as discussed in
  Ref.\cite{DeFelice:2010aj}.} initial conditions are written in terms
of the initial values of the slow-roll parameters. The initial
slow-roll parameters are chosen as $\epsilon(t_i) =-\dot{H}/{H^2}
\approx 7\times 10^{-3}$ and $\eta(t_i) =\ddot{H}/(H\dot{H}) \approx
0$. The other initial conditions are as:
\begin{eqnarray}
  a(t_i)&=&1\,,\\
  \dot{a}(t_i)&=&H(t_i)a(t_i)\,,\\
  \ddot{a}(t_i)&=&\frac{(1-\epsilon)\dot{a}(t_i)^2}{a(t_i)}\,,\\
  \dddot{a}(t_i)&=&\frac{(1-\eta\epsilon-3\epsilon) \dot{a}(t_i)^3}{a(t_i)^2}\,.
\end{eqnarray}
Although the initial conditions for inflation in the present section
are written in terms of the slow-roll parameters $\epsilon$ and $\eta$
at initial time we do not evolve $\epsilon$ or $\eta$ with time or use
the slow-roll parameters in any other place in our calculation. The
calculation do not use slow-roll approximation and the results we
present in this subsection are exact results.

To get a numerical solution, we plug the solution
$x(a,\dot{a},\ddot{a})$ into the dynamical equation,
Eq.~(\ref{hdot3}). The resulting dynamical equation for $a(t)$ is a
fourth order ordinary differential equation for the case of quadratic
$f(R)$ gravity. Looking at the structure of the roots of the cubic
equation, presented in the initial part of the present section, it is
seen that in the present case the coefficient $A_1>0$ and consequently
there will be only one real root of the anisotropy factor $x(t)$. This
root continuously tends to zero as $b$ tends to zero.

We have plotted the numerical results showing the growth of the Hubble
parameter and $x$, for small and large relative initial anisotropy, in
logarithmic scales in Fig.~\ref{fig1}. Small anisotropy, $x_{\rm
  small}$, corresponds to $b=10^{-10}$ and relatively large
anisotropy, $x_{\rm large}$, corresponds to $b=10^{-5}$. For both
small and large anisotropies the scale-factor and the Hubble parameter
are approximately the same showing that the overall inflating nature
of the system does not depend upon the  initial anisotropy
present in the system. The inflationary nature of the present system
is shown by the near constant value of $H$ in Fig.~\ref{fig1}.
Fig.~\ref{fig2} shows the growth of the anisotropy factor $x^2/H^2$
in the two cases corresponding to the two $b$ values as discussed
above. This plot clearly shows that anisotropy rapidly dies in
quadratic gravity inflation. We have numerically verified that
anisotropy gets wiped out after the first two or three e-folds of
inflation and consequently suppression of anisotropy happens very
efficiently in quadratic inflation. Here it must be noted that we
cannot arbitrarily increase $b$ as when $b>10^{-4}$ the consistency
condition in Eq.~(\ref{anisb}) is violated and the system does not
inflate any more.  Our formalism shows both analytically and
numerically that Starobinsky inflation is safe from initial
anisotropy.
\subsection{Contraction in toy model of quadratic bounce}
\label{plc}

In this subsection we discuss anisotropic contraction phase in a
simple and partly unstable model, guided by quadratic gravity. The
presentation in this subsection is more like a toy model analysis
which shows the complexities of anisotropic contraction in polynomial
$f(R)$ gravity models. In general the solution of Eq.~(\ref{ddotbeqn})
becomes a polynomial equation in $x$ and for higher polynomial orders
(compared to quadratic order) the algebraic equations do not yield
analytic solutions. The quadratic gravity bounce model, where
$\alpha<0$ is the simplest polynomial bounce model, where the
intricacies of anisotropy generation during the contraction phase can
be semi-analytically shown. 

The issue of anisotropy generation during a contraction phase is very
important as anisotropy may get enhanced during this phase as it
happens in GR based models of cosmological bounce. We want to see how
anisotropy grows in Bianchi-I models in the contraction phase in
quadratic $f(R)$ gravity. In the present case we will assume the
existence of hydrodynamic matter and $\alpha<0$ as these conditions
are required for a subsequent bounce\cite{Paul:2014cxa}. Before we
proceed we will like to make some remarks related to the choice of the
sign of $\alpha$. The negative sign of $\alpha$ implies that
$f^\prime(R)$ is not always positive. We can choose our dynamical
system to be such that it satisfies $f^\prime(R)>0$ for some range of
$R$, as done in the present paper. The importance of negative $\alpha$
quadratic model is that one can have a cosmological bounce in this
restricted regime of $R$, where $R<1/(2|\alpha|)$ for stability. There
is another source of instability in the present case, related to the
negative sign of $\alpha$.  In such models $f^{\prime \prime}<0$ which
may lead to instabilities first proposed by Dolgov and
Kawasaki\cite{Dolgov:2003px} and later by
V.~Faraoni\cite{Faraoni:2006sy}. In the present model one cannot get
rid of Dolgov and Kawasaki instability\footnote{In
  Ref.~\cite{Bhattacharya:2015nda} it was explicitly shown that
  polynomial $f(R)$ gravity theories which accommodate bouncing
  solutions cannot satisfy $f^\prime>0$ and $f^{\prime\prime}>0$ for
  all values of Ricci scalar.}, consequently in light of the stability
issues we will like to interpret the present model of bounce as a toy
model whose sole purpose is to describe the nonlinear growth of
anisotropy. In the a later section we will apply or formalism in a
stable gravitational model.

In GR it is known that anisotropy suppression during contraction phase
requires the presence of an ultra-stiff matter component with $\omega
(=P/\rho) >1$.  The presence of an ultra-stiff matter component can
produce a slow contraction phase where preexisting anisotropy is
suppressed\footnote{Sometimes this phase of slow contraction under the
  dominance of a ultra-stiff matter is called the ekpyrotic
  phase\cite{Garfinkle:2008ei}.}.  In the present case we will see
that a power law contraction phase may suppress initial anisotropy in
quadratic $f(R)$ cosmology.

We assume that during the contracting phase $t<0$ and bounce occurs at
$t=0$.  During the contracting phase the scale-factor decreases as
\begin{eqnarray}
a(t) \propto (-t)^n\,,\,\,{\rm where}\,\,\,0<n<1\,,
\label{csf}
\end{eqnarray}
and consequently
$$H=\frac{n}{t}\,.$$ From physical considerations one can choose $\alpha
= -10^{12}$ in Planck units\cite{Paul:2014cxa}.  Eliminating $\rho$ in
Eq.~(\ref{hdot3}) by using Eq.~(\ref{hsquare}) we get,
\begin{eqnarray}
  \omega = \frac{(4\dot{H} + 6H^2 + x^2)f^\prime + 2 \kappa P_{\rm curv}}
         {(x^2 - 6H^2)f^\prime + 2\kappa \rho_{\rm curv}}\,.
\label{omexp}
\end{eqnarray}
In the present case the above equation yields the equation of state
for the barotropic matter when one specifies the particular nature of
the scale-factor.

Determining the form of the time evolution of anisotropy factor
reduces to finding the root(s) of Eq.~(\ref{dotb}). One can have
various phases of anisotropy development during a cosmological
evolution depending upon the roots of
\begin{figure}[!t]
\centering
\includegraphics[scale=.9]{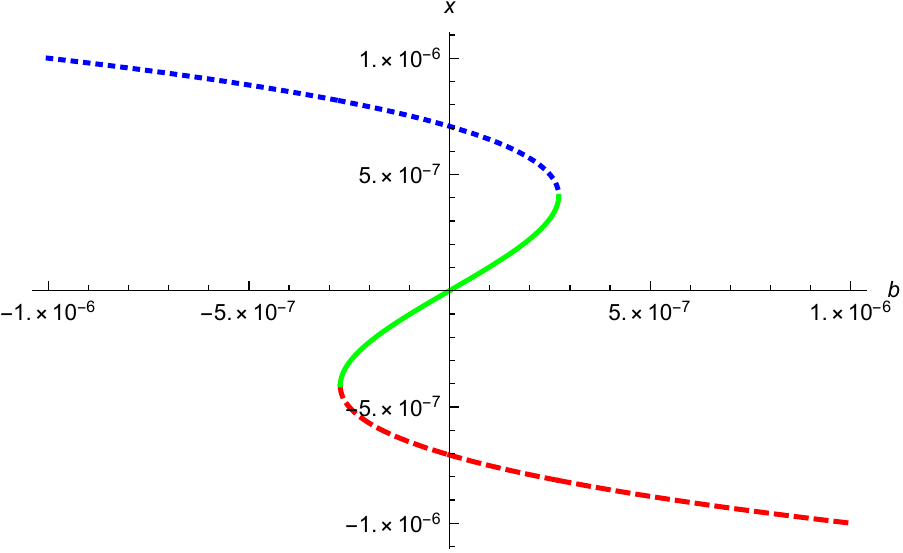}
\caption{Plot of initial anisotropy $x$ with respect to parameter $b$
  at $t=-10^{10}$. The dashed and the continuous parts correspond to
  the (second and third) roots with the minus/plus signs after $\pi$
  in the second line on the right hand side of
  Eq.~(\ref{trigroot}). The dotted part corresponds to the (first)
  root on the first line on the right hand side of
  Eq.~(\ref{trigroot}). All the three roots are real near $b=0$.}
\label{f:3roots}
\end{figure}
Eq.~(\ref{dotb}). In this paper we will particularly focus on the
contracting phase of the universe leading to a cosmic bounce.  We
present the results for the popular quadratic $f(R)$ model which
actually accommodates a cosmological bounce \cite{Paul:2014cxa},
\cite{Carloni:2005ii}, \cite{Bamba:2013fha}, \cite{Barragan:2010qb}.
The nature of the anisotropic contraction phase predicted in this
model will give a glimpse of the interesting effects of $f(R)$ models
of anisotropic contraction in the Bianchi-I spacetime.  The plot in
Fig.~\ref{f:3roots} shows the nature of the roots at time $t=-10^{10}$
in Planck units.  The time period of contraction is chosen in such a
way that all the constraints as $f^\prime(R)>0$ and $\rho>0$ are
maintained during this phase of contraction. As the power law
contraction can never lead to a bounce the constraints compel us to
terminate the power law contraction process some time before the
bounce and in this paper we use the time interval $-10^{10}\le t \le
-10^7$.  The scale-factor during this time is assumed to be
$a(t)=(-t/10^{10})^n$ such that $a(t=-10^{10})=1$.  The nature of the
roots show that below a certain $b$ value and above a certain $b$
value there is only one real root. Near $b=0$ the system admits three
real roots of $x(t)$. We have verified that the nature of the root
structure, as specified in Fig.~\ref{f:3roots}, does feebly depend
upon $n$ in the interval $0 \le n \le 1$. The plot in
Fig.~\ref{f:3roots} shows three branches in three colors. The middle
green (continuous line) branch smoothly matches to the blue (dotted) branch
above and the red (dashed) one below.  The continuous branch specifies a root of
Eq.~(\ref{cubice}) which is real near $b=0$ and gives rise to small
values of anisotropy factor $x_0=x(t=-10^{10})$ initially. 
\begin{figure}[!t]
\begin{minipage}[b]{0.6\linewidth}  
\centering
\includegraphics[scale=.6]{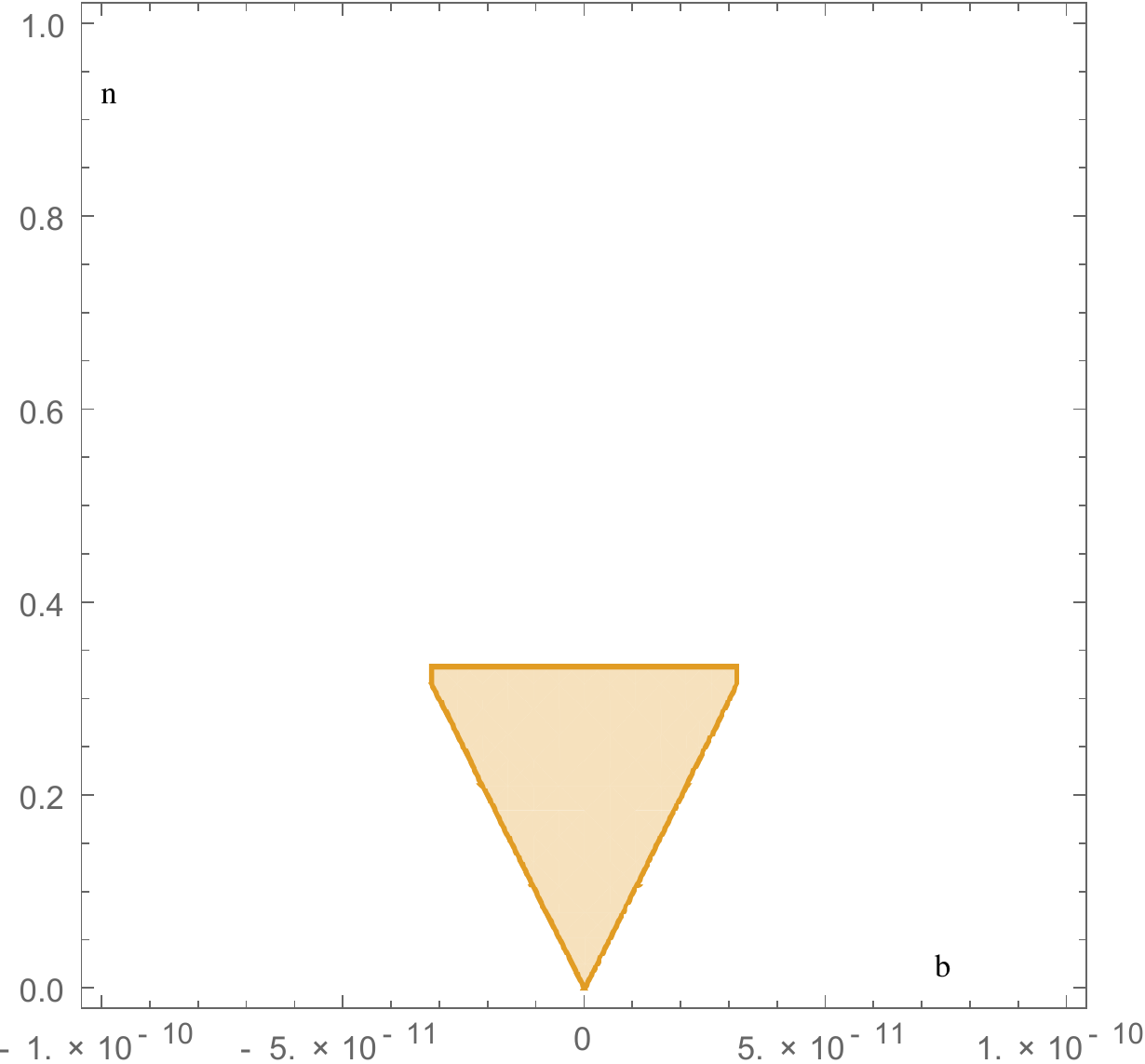}
\caption{Region in the $b-n$ plane giving rise to decreasing
  $x^2/H^2$. Here the abscissa is specified by the $b$ values
  and the ordinate is specified by $n$ values.} 
\label{f:bns}
\end{minipage}
\hspace{0.2cm}
\begin{minipage}[b]{0.6\linewidth}
  \includegraphics[scale=.6]{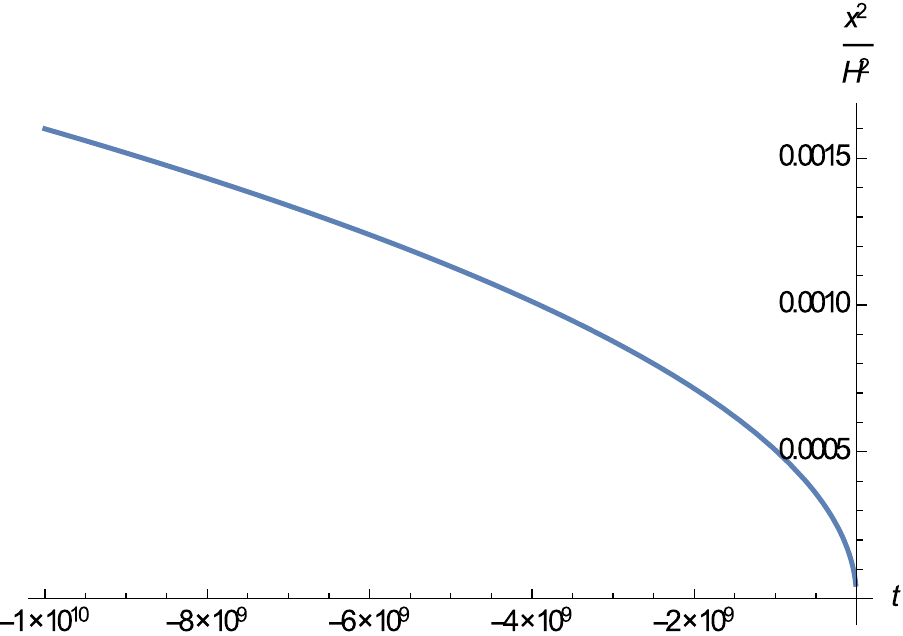}
\caption{Decrease of anisotropy in time when $b,n$ lies in the shaded
  region in Fig.~\ref{f:bns}. The specific $b,n$ value chosen for the
  plot is given in the text.}
\label{f:hxs}
\end{minipage}
\end{figure}
The dashed and dotted branches specify the other roots which are large
for regions near $b=0$. The connection of the three regions in the
figure with the roots in Eq.~(\ref{trigroot}) are specified in the
caption of Fig.~\ref{f:3roots}. As time evolves the nature of the plot in
Fig.~\ref{f:3roots} changes but the general structure of the plot
always remains qualitatively similar as the one plotted at the initial
time.

The dynamics of anisotropy growth depends upon the parameters $b$ and
$n$. We can specify the region in the $b-n$ plane which gives rise to
decreasing anisotropy. The plot in Fig.~\ref{f:bns} shows such a
region in the $b-n$ plane. The plot is done at $t=-10^{10}$, the
initial time, when the region is most constrained. In Fig.~\ref{f:hxs}
we show how $x^2/H^2$ varies in time if one uses any value of $b,n$ in
the shaded region in Fig.\ref{f:bns}. In particular we have chosen
$b=10^{-12}$ and $n=1/4$.  The above information shows that for some
parameter values a power law contraction can indeed suppress small
anisotropy in quadratic gravity. Physically anisotropy suppression for
some regions in the $b-n$ plane in the contracting phase is not 
surprising as both $x$ and $H$ do increase in time during contraction
when $b$ and $n$ belongs to the shaded region in Fig.~\ref{f:bns} (as
expected) but $H$ increases more than $x$ in time and as a consequence
$x^2/H^2$ decreases with time. For some
parameter space $H$ can grow faster than $x$ in time, when anisotropy factor
decreases, and for other parameter values $x$ increases more than $H$
in time making the contracting universe completely anisotropic. 
\section{Anisotropy growth in exponential gravity}
\label{expgrv}

Recently it has been shown\cite{Bari:2018aac}  that one can get bouncing solutions
and expanding universe solutions in a unstable de-Sitter point in
exponential gravity where
\begin{eqnarray}
f(R)=\frac{1}{\alpha}e^{\alpha R}\,,\,\,\,\,\alpha>0\,,
\label{expf}
\end{eqnarray}
where $\alpha$ is a dimensional, real constant. In the present case as
$\alpha>0$ we have $f^\prime(R)>0$ and $f^{\prime \prime}(R)>0$ and the
theory remains stable for all values of $R$.  It was shown in
Ref.\cite{Bari:2018aac} that exponential gravity do admit some exact
solutions. One exact solution is a bouncing solution and another one
is an expanding universe solution with constant Hubble parameter which
takes place at a de-Sitter point. As we know two exact solutions in
exponential gravity we can investigate about the growth of anisotropy
in these two cases.

In the present case the solution of anisotropy factor is,  
\begin{eqnarray}
  x=\frac{b}{f^\prime(R)}=\frac{b}{a^3 e^{\alpha R}} =
  \frac{b}{a^3 e^{6\alpha (\dot{H} + 2H^2)}} e^{-\alpha x^2}\,.
\label{xexp}
\end{eqnarray}
This is a transcendental equation in $x(t)$. The form of the above
equation also shows that there will be only one real solution at any
given time, given graphically by the intersection of a straight line
$y=x$ and a Gaussian $y=\frac{b}{a^3 e^{6\alpha (\dot{H} + 2H^2)}}
e^{-\alpha x^2}$. The small anisotropy solution can be obtained
analytically. If we want to see how small anisotropy, defined by all
the $x$ values which satisfy $x^2 \ll H^2$, develop we may approximate
the last equation as:
\begin{eqnarray}
  x \sim \frac{b}{a^3 e^{\alpha R_{\rm iso}}}\,.
\label{sxexp}
\end{eqnarray}
where $R_{\rm iso} \equiv 6(\dot{H} + 2H^2)$ and $a$ is the average
scale-factor.  We discuss the evolution of small anisotropy for the
two exact solutions of exponential gravity which was extensively
discussed in \cite{Bari:2018aac}.
\subsection{Bouncing solution}

Exponential gravity has an exact bouncing solution, where the
scale-factor is given by $a(t)=e^{At^2}$ where $A$ is a real
constant. Bounce happens in the presence of matter at $t=0$, and the
conditions for an exact solution requires $\alpha A=1/48$ and the
equation of state of matter $\omega=-4/3$. In the present case $R_{\rm
  iso}=12A(1+4At^2)$ and consequently for small anisotropy we must
have
\begin{eqnarray}
x(t) = \frac{b}{e^{1/4}} e^{-4At^2}\,,
\label{sxb}
\end{eqnarray}
which shows that how the anisotropy factor changes with time. The real
indicator of anisotropy is the ratio $x^2/H^2$ and in our present case
\begin{eqnarray}
\frac{x^2}{H^2} = \frac{b^2}{4A^2 \sqrt{e}} \frac{e^{-8At^2}}{t^2}\,.
\label{x2h2c}
\end{eqnarray}
A small anisotropy ratio at $t \to -\infty$ remains smaller than one
for some time but then after some finite time $x^2 \sim H^2$ and
cosmic dynamics is guided by the anisotropy factor leading to an
instability. From our simple analysis we see that the specific
bouncing scenario presented in this section is unstable under small
values of anisotropy.
\subsection{Expansion with constant Hubble parameter at the de-Sitter point}

Exponential gravity has another exact, constant Hubble parameter
solution at a de-Sitter point where $R_{\rm iso}=2/\alpha$. The
scale-factor of the universe at the de-Sitter point is $a(t)=e^{H t}$
and this is a vacuum solution when $H^2=1/(6\alpha)$ is satisfied. In
the present case small anisotropy grows as
\begin{eqnarray}
x(t) = \frac{b}{e^2} e^{-3Ht}\,.
\label{sxe}
\end{eqnarray}
and consequently
\begin{eqnarray}
\frac{x^2}{H^2}=\frac{6\alpha b^2}{e^4}e^{-\sqrt{(6/\alpha)} t}\,.
\label{x2h2e}
\end{eqnarray}
In this case we see that small anisotropy decreases with time. This
analysis is not complete as we do not know how large anisotropy
behaves in these situations. To tackle the question of large
anisotropy one has to purely rely on numerical methods.
\section{Conclusion}

This paper presents the general results for anisotropic cosmological
development in Bianchi-I model in metric $f(R)$ gravity. The initial
part of the paper develops the formalism which can be used to track
cosmological development in homogeneous and anisotropic Bianchi-I
model. The formalism developed is dynamically complete and can predict
the development of all the relevant cosmological and fluid parameters
in cosmological time. The methods developed in this paper can be
applied to expanding as well as contracting phase of the universe.  As
anisotropy reduces in the expanding phase in GR it does not mean that
this rule will be generally followed in $f(R)$ cosmology as the
equation predicting anisotropy growth is non-linear in nature and may
have surprises in store. Our preliminary calculations predicts that
inflation in quadratic $f(R)$ cosmology, in Bianchi-I spacetime,
indeed suppresses anisotropy. The results related to inflationary
models in anisotropic spacetimes in $f(R)$ theory are presented in
full details in the present paper. We first show that for Bianchi-I
type models one can analytically prove that anisotropy fades away in
quadratic gravity inflation. We numerically show the validity of our
analytic proof.

As anisotropy development demands special attention in the contracting
phase in cosmological models based on GR our aim was to see how the
problem translates into $f(R)$ cosmology.  In this article we tried to
verify whether anisotropy subsides in the $f(R)$ theory driven
contraction phase. The result we obtain is complex and opens up new
areas of research. We have chosen quadratic $f(R)$ theory to
illustrate our results as in this case most of the calculations can be
done analytically although the bouncing scenario is gravitationally
unstable. For any other higher order polynomial $f(R)$ one has to use
numerical methods to determine the solutions of the differential
equation predicting anisotropy dynamics. Our work shows the
qualitative nature of the cosmological system, guided by quadratic
gravity, undergoing anisotropic contraction and we expect
qualitatively similar but quantitatively much more formidable results
for other complicated, gravitationally stable polynomial $f(R)$
cosmologies. Even in the case of quadratic gravity the various results
coming out from our formalism is non-trivial. We have pointed out that
even when we restrict the cosmological dynamics by enforcing
conditions as $f^\prime >0$ and $\rho>0$ there appears various regions
in the $n,b$ plane which gives rise to different kind of anisotropy
growth. For some possible cosmological evolutions we have shown that
anisotropy reduces with time.  There exists other possibilities where
anisotropy increases with time during the contraction phase in
quadratic gravity.

As quadratic $f(R)$ theory cosmological bounce is more like a toy
model because in this case the cosmological dynamics is unstable we
have tried to show the applicability of our result in stable
exponential gravity model which admits an exact bouncing solution. In
this case we have not presented a general result but has focussed on
small anisotropy growth. Our result shows that the exact bouncing
solution in exponential gravity model is unstable and consequently the
cosmological system will tend towards an instability in the
contraction phase. We have also showed that small anisotropy subsides
in the expansion phase at the de-Sitter point in exponential gravity.

The present paper shows that the issue about anisotropy in Bianchi-I
spacetimes in metric $f(R)$ gravity is a nonlinear problem which may
lead to very complex conditions in contracting regions of a bouncing
model. For polynomial gravity theories the cosmological contraction
process is much involved and requires full numerical simulation to
find out meaningful results. For expanding cosmologies our theory has
given expected results, the amount of anisotropy goes down with
expansion. But whether anisotropy will reduce for all kinds of
expansion processes requires a more general proof and we hope we will
able to show more general and formal work in these lines in the near future.  

\end{document}